\documentclass[prb,aps,showpacs,floatfix]{revtex4}
\usepackage{graphicx}
 
\begin{document}

\title{Interplay between magnetism and superconductivity in URhGe}
\author{V. P. Mineev} 
\affiliation{Commissariat \'a l'Energie
Atomique, DSM/DRFMC/SPSMS 38054 Grenoble, France}

\date{June 9, 2005}

\begin{abstract}
The magnetization rotation transition occurs in the itinerant
ferromagnet $URhGe$ when the field about 12T is applied in direction
perpendicular to spontaneous magnetization in the plane of the
smallest magnetic anisotropy energy.The transition is accompanied by the
maximum of resistivity in the normal state and by reentrance of
superconductivity at lower temperatures in the field interval between 8 and
13 Tesla [F.Levy et al, Science, to be published (2005)].

We discuss the magnetization orientation transition and the modification of
triplet pairing superconducting state coexisting with ferromagnetism
up to the fields about 2 Tesla and then reappearing in the broad
vicinity of the transition.
    
The nonsymmorphic space group crystal symmetry of ferromagnetic $URhGe$ allows
existance of antiferromagnetic ordering of magnetic moments of pairs of
uranium atoms along $a$ axis.  We show that the amplitude of this weak
antiferromagnetic ordering increases below the phase transition into
superconducting state due to Cooper pairs spontaneous magnetism.

\end{abstract}

\maketitle
    
\bigskip 

\section{Introduction}

The new class of superconducting materials $UGe_{2}¥$,
\cite{1}
$URhGe$ \cite{2}, and $UIr$ \cite{3} has been revealed recently where
the superconducting state coexists with itinerant ferromagnetic
ordered state.  The large band splitting and the high low temperature
value of upper critical field \cite{4,5} in uranium ferromagnetic
superconductors point out that here we deal with Cooper pairing in the
triplet state which earlier has been attributed with evidence only to
superfluid phases of liquid Helium-3 \cite{6}.  The discovery has
brought into the open the intriguing possibility of unconventional
mechanism of pairing or magnetically mediated superconductivity which
is now under intensive investigation (see for instance \cite{7} and
references therein).  At the same time essential progress was achieved
in general phenomenological description of triplet superconducting
states in ferromagnetic metals \cite{8,9,10}.

The superconductivity of
itinerant ferromagnets presents the particular example of multiband
superconductivity \cite{11,12}.  Its another manifestation is recently
found in conventional two band superconductor $MgB_{2}¥$ \cite{13}. 
The multiband effects also can be important in superconductors without
inversion center $-$ another hot point of up-to-date condensed matter
physics (see \cite{12} and references therein).

The coming to light of superconductivity in any new material give rise the
problem of determination of the type of the superconducting state.  In
particular it is always important to know: (i) what kind of superconductivity
conventional or nonconventional we deal with (it means: is the symmetry
of the order parameter lower than the symmetry of the crystal in the
normal state or it is not \cite{14}, 
(ii) is it singlet or triplet type of the Cooper
pairing, (iii) are there the nodes in the superconducting
quasiparticle spectrum, and if, yes, what kind of the nodes we deal
with symmetry nodes or with occasional nodes, (iv) is the
superconducting state magnetic or not, another words, does it possess
spontaneous magnetic moment or it does not. 

In the present paper we discuss the magnetic properties of
ferromagnetic superconductor $URhGe$: first, the quite recently
discovered \cite{15} metamagnetic transition at the magnetic field about 12
Tesla directed perpendicular to spontaneous magnetization accompanied by the
maximum of resistivity in the normal state and by reentrance of
superconductivity at lower temperatures in the broad vicinity of this
transition between 8 and 13 Tesla.  We shall follow up the
modifications of superconducting state in the ferromagnetic $URhGe$
under magnetic field perpendicular to spontaneous magnetization and
argue that the reentrance of the superconductivity under magnetic
field is compatible with triplet Cooper pairing in this material.

Secondly, we shall describe the interplay between the nonunitary
triplet superconductivity and the weak antiferromagnetism  allowed
by symmetry in this ferromagnet.  The effect of stimulation of weak
antiferromagnetism by the Cooper pairs magnetic moment will be
demonstrated.  The latter reveals the new possibility of direct
experimental determination of spontaneous magnetism  in $URhGe$
superconducting state.

The plan of the paper is as follows. In the next Section we will describe the
metamagnetic transition.  Then, the overview of the triplet
superconducting states in orthorombic ferromagnets will be given,
followed by the description of modifications acquired by superconducting state
under magnetic field.  In the last Section we present the symmetry
analysis of the interplay between the ferromagnetism,
antiferromagnetism and superconductivity in $URh Ge$.

\section{Magnetic orientation transition}

$URhGe$ has the orthorombic crystal symmetry with ferromagnetic moment 
directed in the direction of $c$ crystallography axis.  The experimental 
observations of field dependences of magnetization along different
crystallogphic directions \cite{16} as well as numerical calculations
of magnetic anisotropy energy \cite{17} yield the c-axis to be the
easy magnetization axis and the $a$ axis the hard magnetization axis. 
The magnetic anisotropy energy in $a-c$ plane is more than the four
times larger than in $b-c$ plane meaning that $URhGe$ is the $b-c$
easy magnetization plane ferromagnet.  Following Ref.[17] one can
write the $b-c$ plane anisotropy energy as
\begin{equation}
E_{anis}¥(\theta, H)=\alpha \sin^{2}¥ \theta +(\beta/4)\sin^{2}¥2\theta -M H
\sin \theta,
\label{e1}
\end{equation}
where $\theta$ is the angle between $c$ axis and total magnetization
in $b-c$ plane, magnetic field $H$ directed
along $b$ axis.  The anisotropy energy may be
considered as a part of the total Landau energy of ferromagnet in
magnetic field
\begin{equation}
F=\alpha_{z}¥(T)M_{z}¥^{2}¥+ \alpha_{y}¥ M_{y}¥^{2}¥+
\beta_{z}¥M_{z}¥^{4}¥+\beta_{y}¥M_{y}¥^{4}¥+2\beta_{yz}¥
M_{z}¥^{2}¥M_{y}¥^{2}¥-M_{y}¥ H.
\label{ea}
\end{equation}
Here the $y,z$ are directions of the spin axes pinned to $(b,c)$
crystallographic directions correspondingly.
At the temperatures below the Curie
temperature $\alpha_{z}¥(T)<0$ and in the absence of magnetic field
the $z$ component of magnetization has a finite value.  The magnetic
field creates the magnetization along its direction but decreases the
magnetization parallel to $c$.  This process is abruptly finished at
some field then the $M_{c}¥$ drops to zero.  Another words, in a
ferromagnet with magnetization directed along $c$ axis under magnetic
field directed along $b$ axis there is the first order type transition
between the states with magnetization projections $(M_{y0}, ¥M_{z0}¥)$
and $(\tilde {M}_{y0},~0)$.  To prove this statement one must
investigate the evolution of minima of the free energy depending on two
projections of magnetization.  This is strightforward but a bit
cumbersome problem.  Instead, in assumption of constancy of
magnetization modulos we do much easier investigation of anisotropy
energy (\ref{e1}) depending of just one angular variable.

The anisotropy energy at $H=0$ has two minima: absolute at
$\theta_{1}¥=0$ and metastable at $\theta_{2}¥=\pi/2$ and one maximum in
between of them given by $\sin^{2}¥\theta_{m}¥=(\alpha+\beta)/2\beta$
that is $\theta_{m}¥\approx 60^{\circ}¥$ if we take the numerical
values of coefficients $\alpha+\beta\approx 4.4 meV$ and $\beta=2.9
meV$  found in Ref.[17].  The values of anisotropy energy in both minima decrease with
increasing field but the metastable minimum $E_{anis}¥(\pi/2, H)$
drops faster and at some field becomes deeper.  Still at this point
between two minima there is the maximum: It is easy to check by direct
calculation that $\partial^{2}¥E_{anis}¥(\theta,
H)/\partial\theta^{2}¥|_{\theta=\pi/2}¥ $ is positive at arbitrary
magnetic field and the values of $\alpha$ and $\beta$ parameters
pointed out above.  Hence we have the first order type transition from
the state with finite $M_{z}¥$ component of magnetization to the state
there this component is absent.

The crystal symmetry is changed with magnetic field increasing.  First, at
zero field the magnetic symmetry group of the orthorombic crystal with
magnetization oriented along $c$ axis is
\begin{equation}
D_{2}(C_{2}^{z})=
(E, C_{2}^{z}, KC_{2}^{x},
KC_{2}^{y})    
\label{e3} 
\end{equation}
where $K$ is the time reversal operation.
Then at intermediate fields the magnetization has both $M_{b}¥, M_{c}¥$
components and the crystal symmetry is decreased to monoclinic
\begin{equation}
C_{2}¥^{x}¥= (E, KC_{2}^{x}).
\label{e4} 
\end{equation}
Finally after the first order type transition the orthorombic
symmetry is recreated but with magnetization directed along $b$
direction
\begin{equation}
D_{2}(C_{2}^{y})= (E, KC_{2}^{z}, KC_{2}^{x}, C_{2}^{y})
\label{e5} 
\end{equation}

\section{Superconducting states in the orthorombic ferromagnet
with triplet pairing}

The symmetry
description of all possible superconducting states in orthorombic
ferromagnets was given in \cite{9,11,12}.  We shortly repeat here the
main points of this description.
In an itinerant ferromagnetic metal 
the internal exchange field lifts the Kramers degeneracy of the
electronic states.  The electrons with spin "up" fill the states in
some bands and the electrons with spin "down" occupate the states in
other bands.  Hence we have the specific example of multiband metal
with states in each band filled by electrons with only one spin
direction. Let us discuss for simplicity the two-band ferromagnet. 
If there is some pairing interaction, one can discuss intraband or
spin "up" - spin "up" ( spin "down"-spin "down") pairing of electrons,
as well as interband or spin "up"-spin "down" pairing.  In general the
Fermi surfaces of spin up and spin down bands are situated in
different places of the reciprocal space and have the different shape. 
That is why pairing of electrons from the different bands occurs just
in the case of nesting of some peaces of the corresponding Fermi
surfaces.  In such the situation, similar to SDW or CDW ordering, the
superconducting ordering is formed by Cooper pairs condensate with
finite momentum known as Fulde-Ferrel-Larkin-Ovchinnikov state.  We
shall not discuss here this special possibility.  So we neglect by
pairing of electronic states from different bands giving Cooper pairs
with zero spin projection.  Hence, the only superconducting state
should be considered it is the state with triplet pairing and the
order parameter given by
\begin{equation}
{\bf d}^{\Gamma}¥({\bf R},{\bf k})=\frac{1}{2}
[-(\hat{x}+i\hat{y})\Delta_{\uparrow}¥({\bf R},{\bf k})+
(\hat{x}-i\hat{y})\Delta_{\downarrow}¥({\bf R},{\bf k})]
\label{e6}
\end{equation}
Superconducting states ${\bf d}^{\Gamma}¥({\bf R},{\bf k})$ with different
critical temperatures in the ferromagnetic crystals are classified in
accordance with irreducible co-representations $\Gamma$ of the
magnetic group $M$ of crystal~\cite{8,9}.  All the co-representations
in ferromagnets with orthorombic symmetry are one-dimensional. 
However, they obey of multicomponent order parameters determined
through the coordinate dependent pairing amplitudes: one per each band
populated by electrons with spins "up" or "down".  For the two-band
ferromagnet under discussion, they are
\begin{equation} 
\Delta_{\uparrow}¥({\bf R},{\bf k})=-\eta_{1}¥({\bf R})f_{-}¥({\bf
k}),~~~~ \Delta_{\downarrow}¥({\bf R},{\bf k})=\eta_{2}¥({\bf
R})f_{+}¥({\bf k}).
\label{e7}
\end{equation}
The coordinate dependent complex order parameter amplitudes
$\eta_{1}¥({\bf R})$ and $\eta_{2}¥({\bf R})$ 
are not completely independent:
\begin{equation}
\eta_{1}¥({\bf R})=|\eta_{1}¥({\bf R})|e^{i\varphi({\bf R})}¥,
~~~\eta_{2}¥({\bf R})=\pm |\eta_{2}¥({\bf R})| e^{i\varphi({\bf R})}.
\label{e8}
\end{equation}
Being different by their modulos they have the same phase with an
accuracy $\pm \pi$.  The latter property is due to the consistency of
transformation of both parts of the order parameter under the time
reversal.

The general forms of odd functions of momentum directions of pairing
particles on the Fermi
surface $f_{\pm}¥({\bf k})=f_{x}¥({\bf k})\pm if_{y}¥({\bf k})$ for the
different superconducting states in ferromagnets can be found
following the procedure introduced in \cite{9}.  We shall not repeat it here
but just write the order parameter corresponding to "conventional"
superconductivity in a orthorombic ferromagnet with magnetic moment
oriented along $\hat c$ direction.  The symmetry group of such a
crystal is given by eqn.(\ref{e3}). The "conventional" superconducting
state obeys the same symmetry (\ref{e3}) as the normal state and only
the gauge symmetry is broken.  The general form of the order
parameter ${\bf d}({\bf R},{\bf k})$ given by eqns
(\ref{e6})-(\ref{e7}) compatible with symmetry (\ref{e3}) is obtained
by the following choice of the functions $f_{\pm}¥({\bf k})$:
\begin{equation}
f_{\pm}({\bf k})=
k_{x}¥(u_{1}¥\mp u_{4}¥)
+ik_{y}¥(u_{2}¥\pm u_{3}¥),
\label{e9}
\end{equation}
where $u_{1}¥,\ldots $ are real functions of $k_{x}¥^{2}¥,
k_{y}¥^{2}¥, k_{z}¥^{2}¥$.
From the expression for the order parameter one can conlude  that the
only {\it symmetry dictated nodes in quasiparticle spectrum} of
conventional superconducting states in orthorombic ferromagnets are
the nodes lying on the nothern and southern poles of the Fermi surface
$k_{x}¥=k_{y}¥=0$.  Along with superconducting state given by eqn
(\ref{e9}) there is another equivalent superconducting state
transforming as $i{\bf d}^{*}({\bf R},{\bf k})$.  One can prove
\cite{9} that these two states coexist in the same ferromagnetic crystal but
in domains with the opposite direction of magnetization.

All the superconducting states in the orthorombic ferromagnets and in
particular the conventional superconducting state 
are non-unitary and obey the {\it Cooper pair spin momentum}
\begin{equation} 
{\bf S}=i\langle {\bf d}^{*}\times {\bf d}\rangle=
\frac{\hat z}{2}\langle|\Delta_{\uparrow}¥|^{2}¥-
|\Delta_{\downarrow}¥|^{2}¥\rangle ,
\label{e10}
\end{equation}
and {\it Cooper pair angular momentum}
\begin{equation} 
{\bf L}=i\langle {\bf d}^{*}_{\alpha}\left({\bf k}\times
\frac{\partial}{\partial {\bf k}} \right) {\bf d}_{\alpha}\rangle=
\frac{i}{2}\langle \Delta_{\uparrow}¥^{*}\left({\bf k}\times
\frac{\partial}{\partial {\bf k}} \right)\Delta_{\uparrow}¥+
\Delta_{\downarrow }¥^{*}\left({\bf k}\times \frac{\partial}{\partial
{\bf k}} \right)\Delta_{\downarrow } \rangle,
\label{e11}
\end{equation}
where the angular brackets denote the averaging over
${\bf k}$ directions.  As the consequence, the magnetic moment
of ferromagnet changes at the transition to the ferromagnetic
superconducting state \cite{12}.  We shall denote this changement as ${\bf
M}_{s}¥$.  

Certainly, below the phase transition of a ferromagnet to
the superconducting state its  magnetic moment is screened by the
London supercurrents flowing around the surface of the specimen
\cite{14}.  In the case of $UGe_{2}¥$ and $URhGe$ this screening is,
however, uncomplete just becouse the size of ferromagnetic domains
\cite{a} and the London penetration depth Ref.[2] have the same order
of magnitude $\sim 10^{-4}¥cm$.  It is known \cite{b,14} that even  in the
absence of the external field the Abrikosov vortices penetrate into the bulk
ferromagnet if the spontaneous ferromagnetic moment exceeds the lower
critical field $M_{0}¥>H_{c1}¥$.  In presence of domain srtucture this
criterium is modified \cite{c} as follows
$M_{0}¥>H_{c1}¥(\lambda/w)^{2/3}¥$, here $\lambda$ is the London
penetration depth and $w$ is the domain wall thickness.  To operate
with measurable values one can rewrite this inequality as
\begin{equation} 
M_{0}¥>\left (\frac{\xi_{0}¥}{w\kappa^{2}¥}\right )^{2/3}¥H_{c2}¥.
\label{e x}
\end{equation}    
Here
$H_{c2}¥\approx 2T$ is the upper critical field, $\xi_{0}¥\approx
2\cdot 10^{-6}¥cm$ is the coherence length, $\kappa= 50\div 100$ is the
Ginzburg-Landau parameter.  The value of spontaneous magnetization in
$URhGe$ is $10^{2}¥/2~ G$.  Taking the domain wall width as
$w=10^{-7}¥\div 10^{-6}¥cm$ we see that just the opposite inequality
takes place.  Hence in the absence of the external field the domain
structure in $URhGe$ is vortex free.

\section{Field induced superconductivity}

As we already mentioned the magnetization orientation transition is
accompanied at low temperatures \cite{15} (below 0.4 K) by reappearence
of superconductivity.  The phenomenon of magnetic field induced
superconductivity is known more than two decade.  First it was discovered
in pseudoternary molibdenum halcogenides $Eu_{x}¥Sn_{1-x}¥Mo_{6}¥S_{8}¥$
\cite{18}.  The orbital critical field in this materials is likely sufficiently
high due large impurity scattering.  Hence, the critical field value
is mostly controlled by paramagnetic limiting mechanism.  Moreover,
the formation of antiferromagnetic state below 1K ($T_{c}¥=4K$)
completely suppress superconductivity.  But in the fields above 4
Tesla the superconducting state reappears.  The common believe that it is
due to Jaccarino-Peter effect \cite{19} consisting of compensation of the
external applied field by the internal exchange field of magnetic ions
with magnetic moments oriented by high external field.  Similar
phenomenon has been observed in two-dimensional organic superconductor
$\lambda-(BETS)_{2}¥FeCl_{4}¥$ \cite{20,21}.  The very high orbital
field value is maintained here by the field orientation parallel to
conducting layers.  Again due to Jaccarino-Peter mechanism the high
field reentrance of superconductivity occurs.

The quite different situation happens in $URhGe$.  First of all, even
in the low field region the superconducting state exists till about 2
Tesla \cite{5}.  That is about 4 times larger than paramagnetic
limiting field.  The latter means that we deal with triplet
superconductivity and the critical field entirely determined by the
orbital mechanism.  Here we always tell about the field orientation
parallel to $b$ axis that is perpendicular equlibrium direction of
magnetization.  Then, after suppression superconductivity it reappears
at the field about 8 Tesla and persists till about 13 Tesla \cite{15}.

The analysis made in the paper Ref.[5] shows that among the
superconducting states (\ref{e6})-(\ref{e9}) the best fit for the
upper critical fild temperature behavior gives one-band
superconducting state with the order parameter
\begin{equation}
{\bf d}({\bf R},{\bf k})=(\hat{x}+i\hat{y})k_{x}¥\eta({\bf R})
\label{e12}
\end{equation}
As it was described above, under the field influence the magnetization
rotates in the $b-c$ plane from $c$
direction untill it  suddenly proves to be oriented parallel to $b$ axis at
the field value $\approx 12 Tesla$.  We already pointed out that
during the process of magnetization rotation the crystal symmetry is
changed from (\ref{e3}) to (\ref{e5}).  The order parameter form
(\ref{e12}) is compatible with all these symmetry transformations if 
we choose the $\hat y$ axis lying in the plane perpendicular to the total
magnetization direction, such that $\hat x \times \hat y ={\bf M}/M$. 
It is worth noting that similarly one can consider a multiband
superconducting state.

Hence, the order parameter shape is stable in respect to the magnetization
rotation.  This is important observation but it does not explain the
reentrance of superconductivity in the high field region.  Leaving
this problem for the future investigations we only note here that if
the first order type transition is very weak, another words if it is
close to the second order, then in vicinity of it one can expect
appearence of well developed magnetic fluctuations possibly
stimulating of electrons pairing.

\section{Weak antiferromagnetism in superconducting $URhGe$}

The interesting observation has been done in the paper \cite{22} and
discussed in more details in \cite{17}.  The uranium atoms in the
orthorombic unit cell of $URhGe$ form two "pairs" (1,2) and (3,4) called 
$U_{I}¥$ and $U_{II}¥$.  These pairs of atoms can be translate each other by
means of nonprimitive translations, that means the $URhGe$ crystal
lattice is related to nonsymmorphic space group.  It is easy to check
that under the group (\ref{e3}) transformations accompanying by
nonprimitive translations the magnetic moments of uranium atoms behave as
follows
\begin{eqnarray}
&&C_{2}¥^{z}¥~:~~U_{I}¥(M_{x}¥,M_{y}¥,M_{z}¥)\to
U_{II}¥(-M_{x}¥,-M_{y}¥,M_{z}¥)\nonumber\\
&&KC_{2}¥^{x}¥~:~~U_{I}¥(M_{x}¥,M_{y}¥,M_{z}¥)\to
U_{II}¥(-M_{x}¥,M_{y}¥,M_{z}¥)\nonumber\\
&&KC_{2}¥^{y}¥~:~~U_{I}¥(M_{x}¥,M_{y}¥,M_{z}¥)\to
U_{I}¥(M_{x}¥,-M_{y}¥,M_{z}¥)\nonumber\\
&&KC_{2}¥^{y}¥~:~~U_{II}¥(M_{x}¥,M_{y}¥,M_{z}¥)\to
U_{II}¥(M_{x}¥,-M_{y}¥,M_{z}¥)\nonumber\\
\label{e13}
\end{eqnarray}
The symmetry (\ref{e3}) is possible when $M_{y}¥=0$ but magnetization
of pair (1,2) is transformed to the magnetization of pair (3,4) as
$(M_{x}¥,0,M_{z}¥)\to(-M_{x}¥,0,M_{z}¥)$.  Hence along with the
ferromagnetic moment $M_{z}¥$ along $c$ axis, there is possibility of
antiferromagnetic ordering of $M_{x}¥$ component of $U$ pairs (1,2) and(3,4)
along $a$ axis producing noncollinear magnetic ordering in $a-c$ plain
without a further decrease of magnetic symmetry.  This type of
ordering was reported in the paper \cite{22} as result of neutron
powder diffraction experiments.  The authors have found the magnetic
moments of $U$ atoms $0.26 \mu_{B}¥$ canted in $a-c$ plain with angle of 
$\approx\pm 30^{\circ}¥$.  More recent measurements on the polycrystals \cite{2}
do not reproduce the data of Ref.[22].  The reported value of AFM
component in $a-c$ plain has a magnitude smaller than $0.06 \mu_{B}¥$, but 
FM ordered component of $0.37 \mu_{B}¥$ is alighned along $c$ axis.
These data are in good agreement with LSDA calculations Ref.[17]
yielding the AFM component of $0.03 \mu_{B}¥$ and FM component $0.293 \mu_{B}¥$.
To be complete one must mention the recent single-crystal experiments \cite{23}
that report no AFM component, but suggest the collinear ordering
of magnetization confined in $b-c$ plane.  This type of magnetization
direction being away from high symmetry axis means the decrease of
orthorombic symmetry to monoclinic symmetry due to appearence of ferromagnetism.
This is in principle possible but demands from our point the further
experimental confirmation.

So, in $URhGe$ we have FM ordering along $c$ axis and tiny AFM ordering
along $a$ axis of the oppositely directed magnetic moments of $U_{I}¥$
and $U_{II}¥$ pairs of uranium atoms.  The Landau free energy expansion
has the following form
\begin{equation} 
F=\alpha_{z}¥(T) M_{z}¥^{2}¥ + \beta_{z}¥ M_{z}¥^{4}¥ +\alpha_{L}¥L_{x}¥^{2}¥+
\gamma M_{z}¥L_{x}¥,
\label{e14}
\end{equation}
where $L_{x}¥=M_{x}¥(U_{I}¥)-M_{x}¥(U_{II}¥)$ is staggered AF
magnetization.

Here below the Curie temperature $\alpha_{z}¥(T)<0$ and the ferromagnetic
moment has nonzero equilibrium value
$M_{z0}¥^{2}¥\approx-\alpha_{z}¥/2\beta_{z}¥$.  At the same time unlike the
positive value of $\alpha_{L}¥$ the finite AF magnetization appears
\begin{equation} 
L_{x}¥\approx-\frac{\gamma M_{z0}¥}{2\alpha_{L}¥}
\label{e15}
\end{equation}
induced by the ferromagnetism.  The smallness of $L_{z}¥$ is determined by the 
interaction coefficient $\gamma$.  The situation reminds the well
known phenomenon of the weak ferromagnetism \cite{24} allowed by
symmetry in antiferromagnetic crystals and induced by small
relativistic Dzyaloshinskii-Moriya interaction.  Here we have just
the opposite situation: the antiferromagnetic moment allowed by symmetry
in ordered ferromagnetically crystal is induced by small relativistic
interaction.  One can call this phenomena by {\bf weak
antiferromagnetism}.

The tiny value of weak antiferromagnetic ordering in $URhGe$ has not
been revealed experimentally \cite{2}.  It was  pointed out that the
AF component is smaller than $0.06 \mu_{B}¥$. It does not contradict to
the theoretically calculated value \cite{17} yielding
antiferromagnetic component $\approx 0.03 \mu_{B}¥$.  The measurements
of such a small magnetic moments are in frame of experimental resolution. We
note that much smaller values of staggered magnetization 
have been successfully measured in heavy fermionic materials
$URu_{2}¥Si_{2}¥$ \cite{25} and $UPt_{3}¥$ \cite{26}.  The 
experiments \cite{2} has been performed on the polycristalline specimens
at temperatures above $\approx 2K$ that is inside of ferromagnetic
region (the Curie temperature is $T_{C}¥=9.5 K$) but well above the
superconductivity appearence (the critical temperature of
superconducting transition is $T_{s}¥\approx 0.3 K$).  As we pointed
out the superconductivity in $URhGe$ obeys its own ferromagnetic
moment \cite{27} directed parallel to the magnetic moment of ferromagnetic
normal state.  It causes the additional stimulation of the amplitude
of staggered antiferromagnetic moment
\begin{equation} 
L_{x}¥\approx-\frac{\gamma M_{z0}¥+\gamma_{s}¥ M_{s}¥}{2\alpha_{L}}.
\label{e16}
\end{equation}
Hence, below the transition to the superconducting state one can expect the
increasing of staggered antiferromagnetic magnetization.  The experimental 
evidence of this type behavior can serve by the direct verification of
our understanding of specific superconductivity in ferromagnetic
$URhGe$ as nonunitary superconductivity of Cooper pairs with triplet
pairing.

\section{Conclusion}

In conclusion, we have demonstrated an appearence of abrupt
change of magnetization orientation in ferromagnetic $URhGe$  under
magnetic field perpendicular to spontaneous magnetization direction.
Then the form of the superconducting order parameter compatible with all
intermediate magnetic crystal symmetries has been found.

It was shown that particular nonsymmorphic symmetry of ferromagnetic $URhGe$
allows existance of antiferromagnetic ordering of pairs of uranium
atoms along $a$ axis.  The amplitude of this weak antiferromagnetic
order must increase below the phase transition into superconducting
state.  The experimental verification of this is the direct test for
detection of ferromagnetic moment of Cooper pairs.


\begin{thebibliography}{99}
\bibitem{27} The magnetic moment in domains of $URhGe$ is screened
incompletely (see the discussion at the end of Section III).
\end{thebibliography}
\end{document}